\documentclass[11pt]{article}
\usepackage[utf8]{inputenc}

\usepackage{amsmath, amssymb}
\usepackage{booktabs}
\usepackage{hyperref}
\usepackage{multirow}
\usepackage{subcaption}
\usepackage{xcolor}
\usepackage{lineno}
\usepackage{tcolorbox}
\hypersetup{
    colorlinks=true,
    linkcolor=blue,
    citecolor=blue,
    urlcolor=blue
}
\usepackage{url} 
\usepackage{microtype}
\usepackage[english]{babel}
\usepackage{amsmath,amsfonts,graphicx}
\usepackage[square,numbers,sort&compress]{natbib}
\usepackage{fullpage}
\usepackage{lineno}
\usepackage{xcolor}
\usepackage{pdflscape}
\usepackage{hyperref}
\usepackage{multicol,multirow}
\usepackage{authblk}
\usepackage{rotating}
\def\bSig\mathbf{\Sigma}

\title{Gradient-Based Approximate Bayesian Inference with Entropy-Optimized Summary Statistics for Compartmental Models}

\author[1,2,*]{Xiahui Li}
\author[1,2]{Fergus Chadwick} 
\author[1,2]{Ben Swallow}
\affil[1]{School of Mathematics and Statistics, University of St Andrews, KY16 9SS, UK }
\affil[2]{Centre for Research into Ecological and Environmental Modelling, University of St Andrews, KY16 9LZ, UK}

\begin{document}

\label{firstpage}


\maketitle

\begin{abstract}
Recent pandemics have highlighted the critical role of infectious disease models in guiding public health decision-making, driving demand for realistic models that can provide timely answers under uncertainty. Compartmental models are widely used to capture disease dynamics, and advances in data availability, computational resources, and epidemiological understanding have allowed the development of models that incorporate detailed representations of population structure, disease progression, and intervention effects. While these improvements improve model fidelity, they also increase model complexity, leading to high-dimensional parameter spaces, intractable likelihoods, and computational challenges for fitting models to limited surveillance data in real time. Existing likelihood-free methods, such as Approximate Bayesian Computation (ABC) and Bayesian Synthetic Likelihood (BSL), have developed largely independently, each with distinct strengths and limitations. We propose an integrated three-stage framework that synthesizes advances from both likelihood-based and likelihood-free method: (1) ABC-based entropy minimization to identify low-dimensional, approximately orthogonal summary statistics; (2) BSL inference using these optimized summaries to construct tractable Gaussian approximations; and (3) Hamiltonian Monte Carlo sampling for efficient posterior exploration. Through SEIR simulation study and application to the 1978 British boarding school influenza outbreak, we demonstrate that our framework achieves reliable parameter estimation and uncertainty quantification while maintaining computational efficiency.
\end{abstract}

\section{Introduction}
\label{s:intro}

Infectious disease outbreaks present challenges for public health response \citep{thompson2026infectious}. Decision makers need to rapidly answer questions to implement interventions and determine their efficacy: When will the outbreak peak and how severe will it be? Is the outbreak still in its early growth phase, or has transmission begun to slow? Should we implement school closures or lockdowns and, if so, when? The answers to these questions directly inform intervention timing, resource allocation, and communication strategies \citep{cauchemez2008estimating, shearer2020infectious, berger2021rational}. These decisions require reliable estimates of the number of current cases and forecasts of future cases \citep{fischer2016cdc,desai2019real}. Surveillance systems provide time series of reported cases, hospitalizations, deaths, etc, but these data represent indirect and incomplete observations of the underlying transmission process and cannot directly reveal key epidemiological quantities \citep{world2000report,moran2016epidemic}. Consequently, epidemiological models play an important role in combining surveillance data with epidemiological assumptions to infer latent states, quantify uncertainty, and forecast future transmission \citep{grassly2008mathematical,o2010introduction}.

Mechanistic compartmental models are widely used to quantify population-level dynamics and to support outbreak response in multiple stages, including assessing transmissibility, forecasting, and intervention effectiveness evaluation \citep{brauer2008compartmental, osthus2017forecasting}. In this modeling paradigm, individuals in a population transition between compartments that represent health states, with transitions governed by differential equations. Although classical formulations involve only a small number of compartments (e.g., the susceptible-infected-recovered, SIR, model \citep{kermack1927contribution}), modern models often incorporate many more compartments, sometimes involving states to consider age structure, clinical progression, and intervention effects \citep{balabdaoui2020age, chang2021mobility}. Although this flexibility improves realism, it introduces inferential challenges, including high dimensional parameter spaces, latent variables, and uncertainties in the model structure, which complicate the parameter estimation process \citep{swallow2022challenges}.

Bayesian inference provides a powerful statistical framework for tackling these challenges, due to its ability to quantify uncertainty, integrate prior knowledge, and address complex model structures \citep{kypraios2017tutorial, grinsztajn2021bayesian}. Analytically exact solutions are often not possible for even moderately complex Bayesian models, with most practitioners relying on asymptotically exact inference methods, such as Markov chain Monte Carlo (MCMC) \citep{brooks2011handbook}. While the most common implementation of MCMC, the Metropolis-Hastings algorithms \citep{metropolis1953equation, hastings1970monte}, can struggle, advanced techniques like Hamiltonian Monte Carlo (HMC) avoid this by using gradient information to explore parameter space efficiently \citep{neal2012mcmc}. Modern implementations such as Stan's No-U-Turn Sampler (NUTS) automate tuning, making the method accessible without extensive expertise \citep{hoffman2014no, carpenter2017stan}. Despite their advantages, MCMC-based methods are highly dependent on the availability and tractability of likelihood functions, which can be complex or intractable in realistic epidemiological settings. Bayesian inference for models with intractable likelihoods represents a fundamental challenge in modern statistics, arising in diverse scientific domains, from infectious disease epidemiology to ecology, systems biology, and beyond \citep{wilkinson2009stochastic, kypraios2017tutorial}.

Such challenges have motivated the development of a variety of approximate Bayesian inference methods to overcome the intractability of compartmental models: Approximate Bayesian Computation (ABC) and Bayesian Synthetic Likelihood (BSL), which trade off some degree of statistical precision for improved computational feasibility and scalability \citep{wood2010statistical, fasiolo2018extended, price2018bayesian, li2025advances}. Both approaches avoid direct likelihood evaluation, making them attractive for compartmental models where likelihoods are analytically unavailable or prohibitively expensive to compute. They both operate on low-dimensional summary statistics rather than full data, yet their methodological developments have largely progressed in parallel for each approach. For ABC, summary statistics selection is still a challenge, with persistent concerns about approximation bias, information loss, and limited diagnostic tools to assess posterior accuracy \citep{blum2010approximate}. In contrast, research on BSL has focused primarily on reducing the computational cost of estimating the covariance matrix \citep{ong2018likelihood,an2019accelerating,priddle2022efficient}, and relatively little attention has been paid to summary statistic selection within the BSL framework. 

In this paper, we introduce an integrated three-stage Bayesian inference framework to bridge advances in both approximate and asymptotically exact inference methodologies. Our approach combines: (1) ABC-based entropy minimization \citep{nunes2010optimal} to identify low-dimensional, approximately orthogonal subsets of summary statistics that preserve critical information; (2) BSL inference \citep{wood2010statistical, price2018bayesian} constructed using these optimized summaries to improve computational stability and efficiency; and (3) HMC sampling for efficient posterior exploration. We demonstrate the proposed framework using a deterministic SEIR simulation with known ground truth and validate its performance on data from the British boarding school influenza outbreak \citep{center1978influenza}.

The remainder of the paper is organized as follows: Section 2 provides background on summary statistics, ABC, and BSL, comparing the advancement and challenge of both methods. Section 3 introduces our methodological framework. Section 4 benchmarks the performance of our method against simulation studies with known parameters. Section 5 applies the framework to the boarding school influenza outbreak. Section 6 discusses extensions and future directions.

\section{Background and related work}
\label{s:background}

\subsection{Summary statistics for compartmental models}

Outbreak waves have characteristic temporal patterns, as documented in historical epidemics \citep{he2013inferring, camacho2011explaining}. During the early stages of an outbreak, key inferential challenges include estimating transmissibility, quantifying epidemic growth rates, and characterizing transmission heterogeneity \citep{fischer2016cdc, desai2019real}. Traditionally, these questions are addressed using observational data from case-control studies, cohort studies, and household surveys \citep{world2000report, moran2016epidemic}. However, transmission events and locations are rarely observed, surveillance data provide only partial or biased views of the underlying process, and reporting delays and differential healthcare access (limited capacity or willingness to test in certain groups) introduce systematic distortions \citep{heesterbeek2015modeling}. Moreover, the small scale of observational studies may not represent broader epidemic dynamics, challenging reliable estimation of key epidemiological quantities (such as the incubation period or transmission intensity).

Infectious disease dynamics often shares features of chaotic systems, driven by demographic stochasticity, environmental noise, and is partially observed with error. Consequently, epidemic trajectories are highly sensitive to initial conditions such that small changes in the parameters can induce large variations in outcomes, making likelihood-based trajectory matching unstable. Since even identical systems would not reproduce the same outbreak trajectory, inference should instead focus on repeated features of the system \citep{wood2010statistical}, such as exponential growth rate and peak time.

Summary statistics-based inference provides a solution to these challenges. By reducing data to phase-insensitive measures that quantify local dynamic structure and observation distributions, summary statistics preserve the information that is dynamically important in epidemic data while ignoring noise-driven trajectory details \citep{wood2010statistical}.

\subsection{Approximate Bayesian computation}

Approximate Bayesian Computation (ABC) is a likelihood-free inference method that operates on low-dimensional summary statistics rather than full data. It bypasses direct likelihood evaluation by simulating data for the proposed parameter, $\boldsymbol{\theta}$, and accepting those values when the simulated summary statistics $\boldsymbol{s}(\boldsymbol{y}_{\text{sim}})$ are sufficiently close to the observed statistics $\boldsymbol{s}(\boldsymbol{y}_{\text{obs}})$ under a chosen distance metric $d$ and tolerance $\epsilon$ \citep{beaumont2019approximate}. The accuracy of the approximate posterior depends on three choices: summary statistics, distance metric, and tolerance \citep{sisson2018overview, prangle2018summary}. Although extensive developments have improved ABC's performance and scalability \citep{toni2009approximate, jiang2018approximate, retkute2025novel}, the choice of summary statistics remains a major challenge \citep{blum2010approximate}. Existing approaches to summary statistics selection can be broadly categorized into subset selection methods, projection-based methods, and auxiliary-likelihood methods, each have trade-offs \citep{prangle2018summary}. Subset selection methods, for example, preserve interpretability, but there is no guidance on how to choose the candidate set of summaries. Consequently, identifying a set of statistics that is both informative for the parameters of interest and computationally efficient is still an unsolved problem in ABC, with concerns regarding approximation bias and diagnostic tools for posterior accuracy \citep{li2025advances}.

\subsection{Bayesian synthetic likelihood}

An alternative approximate Bayesian inference approach based on summary statistics is the Bayesian Synthetic Likelihood (BSL) method \citep{wood2010statistical, price2018bayesian}. BSL approximates the likelihood of summary statistics by assuming that they follow a multivariate normal distribution $\boldsymbol{s}|\boldsymbol{\theta} \sim \mathcal N(\boldsymbol{\mu}(\boldsymbol{\theta)},\boldsymbol{\Sigma}(\boldsymbol{\theta}))$, with mean $\boldsymbol{\mu}(\boldsymbol{\theta})$ and covariance matrix $\boldsymbol{\Sigma}(\boldsymbol{\theta})$ empirically estimated from model simulations. This forms a synthetic likelihood that approximates the intractable likelihood of the observed summaries $\boldsymbol{s}_{\text{obs}}$. Combining with a prior $\pi(\boldsymbol{\theta})$ produces the synthetic posterior
$$
\pi(\boldsymbol{\theta}|\boldsymbol{s}_{\text{obs}}) \propto \mathcal N(\boldsymbol{s}_{\text{obs}};\hat{\boldsymbol{\mu}}(\boldsymbol{\theta}),\hat{\boldsymbol{\Sigma}}(\boldsymbol{\theta}))\pi(\boldsymbol{\theta}).
$$
In practice, this posterior cannot be evaluated in closed form because $\hat{\boldsymbol{\mu}}(\boldsymbol{\theta}),\hat{\boldsymbol{\Sigma}}(\boldsymbol{\theta})$ must be estimated using Monte Carlo simulation at each $\boldsymbol{\theta}$. As a result, BSL relies on MCMC as a computational bridge, converting likelihood-free inference into a tractable likelihood-based framework \citep{price2018bayesian,frazier2023bayesian}. Most BSL implementations employ random-walk Metropolis-Hastings, such schemes can mix slowly in moderate to high-dimensional parameter spaces. Gradient-based samplers, such as HMC, can offer substantial improvements in efficiency when gradients are available. To date, there has been limited adoption of HMC in BSL methods.

Methodological advances in BSL have focused primarily on mitigating two main challenges: reducing the computational cost of estimating the covariance matrix and the selection of summary statistics that are both informative and compatible with the Gaussian assumption \citep{ong2018likelihood,an2019accelerating,priddle2022efficient}. In contrast, relatively little attention has been devoted to strategies for summary statistic selection within the BSL framework. For a broader overview of Bayesian inference methods in epidemiology, we refer the reader to our recent review \citep{li2025advances}. Here, we focus on the methods most closely related to our proposed framework.

\subsection{Comparison and research gap}

Both ABC and BSL critically depend on summary statistics selection, yet their methodological developments have largely progressed in parallel for each approach. This motivates three key questions: (1) Can recent advances in summary statistics selection for ABC be adapted to improve BSL performance? (2) Which subset of these quantities provides the most information for inference while also maintaining approximate independence? (3) Can summary statistics sets support efficient BSL-based inference with simplified covariance structures? To our knowledge, no existing framework exploits complementary strengths across ABC and BSL. This gap motivates the development of a hybrid inference framework that bridges recent advances in both approximate and exact inference methodologies for epidemic models. 

\section{Inference Methodology}
\label{s:method}

\subsection{Conceptual framework}

\begin{figure}[htbp]
    \centering
    \includegraphics[width = 1 \textwidth]{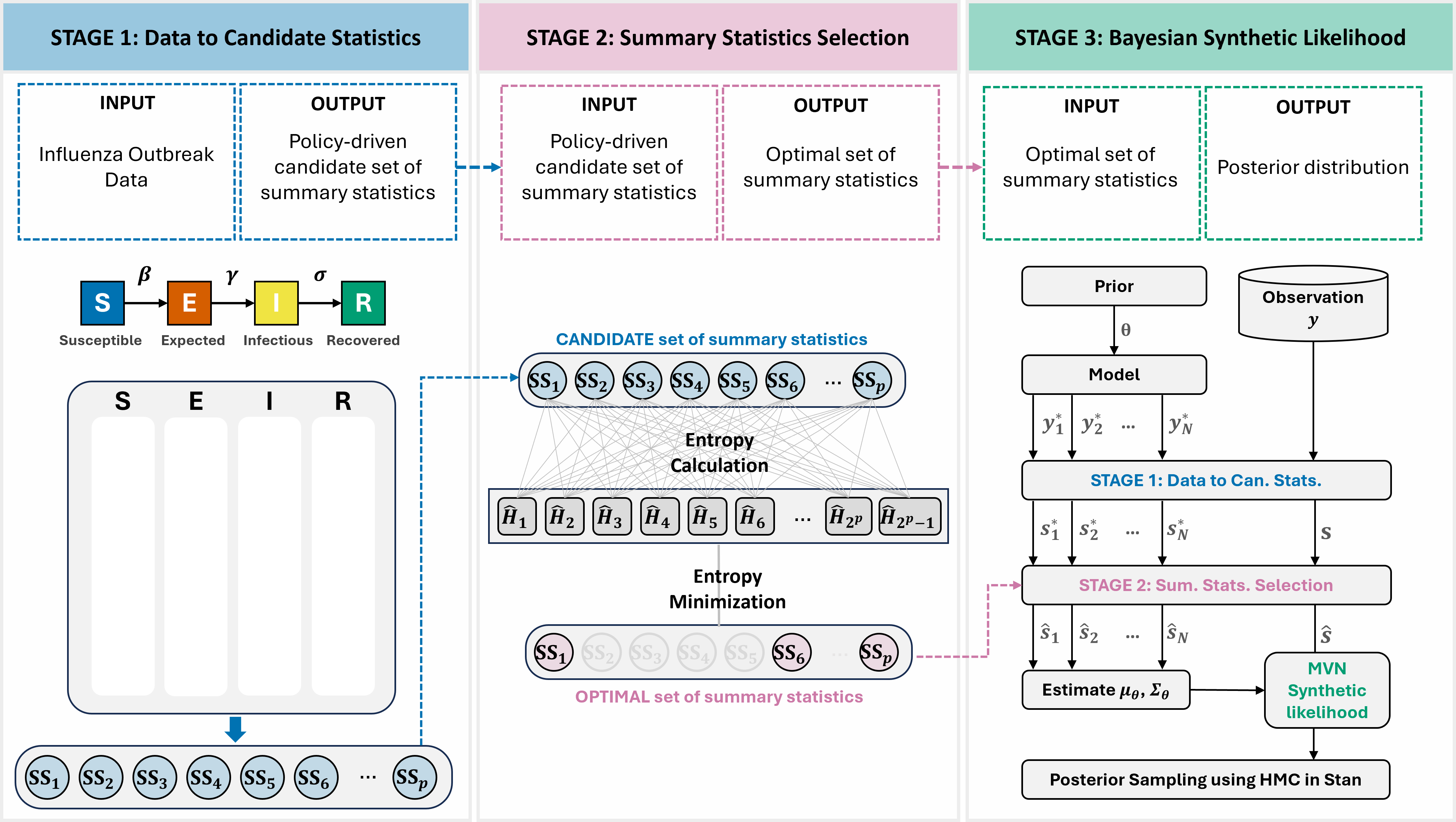}
    \caption{Schematic overview of the proposed three-stage inference framework. In Stage 1, influenza outbreak data are mapped to a candidate pool of policy-driven summary statistics using the compartmental model. In Stage 2, entropy minimization ABC identifies an optimal subset of summary statistics. In Stage 3, the selected statistics are used to construct a multivariate normal synthetic likelihood, allowing posterior sampling through HMC in Stan.}
    \label{figure:workflow}
\end{figure}

We propose a novel hybrid Bayesian inference framework. This approach combines the strengths of ABC, BSL, and HMC to improve computational efficiency and inference accuracy for models with intractable likelihoods. The framework consists of three sequentially integrated components that leverage the complementary strengths of each method. A schematic representation of the proposed method is shown in Figure \ref{figure:workflow}.

\textbf{Stage 1: Policy-driven entropy-based summary statistics selection through ABC.} We begin by constructing a rich set of candidate summary statistics $\boldsymbol{s}(y) = \{s_1(y), s_2(y), ..., s_k(y)\}$ motivated by policy-relevant epidemiological quantities (see \ref{s:policy-driven stats}). To select a low-dimensional subset $\boldsymbol{s}^*(y)$ that preserves maximal information about the parameters $\boldsymbol{\theta}$, we employ an entropy minimization criterion \citep{nunes2010optimal}. Because entropy penalizes overlapping information, highly correlated or conditionally redundant statistics are excluded. This approach combines domain expertise (policy-driven candidate summary statistics ensure epidemiological relevance) with statistical efficiency (entropy minimization preserves information by mitigating redundancy across statistics that inform the same model parameters) to produce a set of mutually exhaustive and sufficient statistics spanning the epidemiological dynamical system.

\textbf{Stage 2: Synthetic Likelihood construction with entropy-selected summaries.} We use these entropy-selected statistics $\boldsymbol{s}^*(y)=(s^*_1(y), s^*_2(y),...,s^*_p(y))$, where $p\leq k$, within BSL framework \citep{wood2010statistical, price2018bayesian}. As in standard BSL, we assume that $ \boldsymbol{s}^*|\boldsymbol{\theta} \sim \mathcal{N}(\boldsymbol{\mu}(\boldsymbol{\theta}),\boldsymbol{\Sigma}(\boldsymbol{\theta})),$ with $\hat{\boldsymbol{\mu}}(\boldsymbol{\theta})$ and $\hat{\boldsymbol{\Sigma}}_i(\boldsymbol{\theta})$ are estimated from $M$ model simulations at fixed $\boldsymbol{\theta}$.

A key advantage of entropy-based selection is that the retained statistics are approximately conditionally independent given $\boldsymbol{\theta}$. This property justifies a diagonal covariance approximation, $ \boldsymbol{\Sigma}(\boldsymbol{\theta}) \approx \text{diag}(\hat{\sigma}^2_1(\boldsymbol{\theta}), ..., \hat{\sigma}^2_p(\boldsymbol{\theta})).$ Under this approximation, the synthetic likelihood thus takes the simplified form:
$$
L_{\text{SL}}(\boldsymbol{\theta};\boldsymbol{s}^*_{\text{obs}})=\mathcal N(\boldsymbol{s}^*_{\text{obs}};\hat{\boldsymbol{\mu}}(\boldsymbol{\theta}), \text{diag}(\hat{\sigma}^2_1(\boldsymbol{\theta}), ..., \hat{\sigma}^2_p(\boldsymbol{\theta})).
$$
The diagonal structure reduces covariance estimation from $O(p^2)$ to $O(p)$ parameters, improving the numerical stability and computational efficiency, particularly as $p$ increases.

\textbf{Stage 3: Efficient posterior sampling via HMC.} Finally, we estimate the posterior distribution $\pi(\boldsymbol{\theta}|\boldsymbol{s}^*_{\text{obs}})$ using HMC sampling implemented in the probabilistic programming language Stan \citep{carpenter2017stan} (see \ref{s:hmc}).

In our methodological framework, each stage builds on the strengths of the previous one. Stage 1 reduces the dimensionality of both time series and summary statistics while retaining the information most relevant for inference, and minimizing computational costs in Stage 2's synthetic likelihood estimation. The resulting summaries provide a tractable and statistically efficient likelihood approximation in Stage 2, enabling Stage 3's gradient-based posterior exploration, often reducing the number of iterations needed for convergence.

\subsection{Policy-driven candidate statistics}
\label{s:policy-driven stats}

An important initial consideration is which summary statistics should we choose to be included in the candidate set of summary statistics. We propose that candidate summary statistics selection should be driven by the policy-relevant questions public health officials need to answer during outbreaks, as these often drive measurable quantities. For an outbreak, decision makers require answers to: when cases will peak, how rapidly cases grow early in the outbreak, and whether the epidemic is in a pre- or post-peak phase \citep{berger2021rational, shearer2020infectious}. Each question naturally suggests the corresponding candidate summary statistics: peak date and peak height address timing and severity; early case growth rate captures transmission intensity; and the overall trajectory shape indicates epidemic phase. These summaries are epidemiologically interpretable and directly inform intervention decisions, including timing of facility closure, isolation capacity planning, and targeted control measures \citep{berger2021rational, cauchemez2008estimating, fischer2016cdc, desai2019real}. 

However, these policy-driven candidates may be redundant (peak day and growth rate contain overlapping information about transmission speed), high-dimensional (tracking multiple trajectory features increases computational cost), or suboptimal for inference (final size may be poorly identified from partial outbreak data observed before the epidemic concludes). Too many statistics increase computational cost and exacerbate the curse of dimensionality in ABC while inflating covariance estimation variance in BSL, while over reduction risks discarding critical information \citep{prangle2018summary}.

\begin{figure}[htbp]
    \centering
    \includegraphics[width = 1 \textwidth]{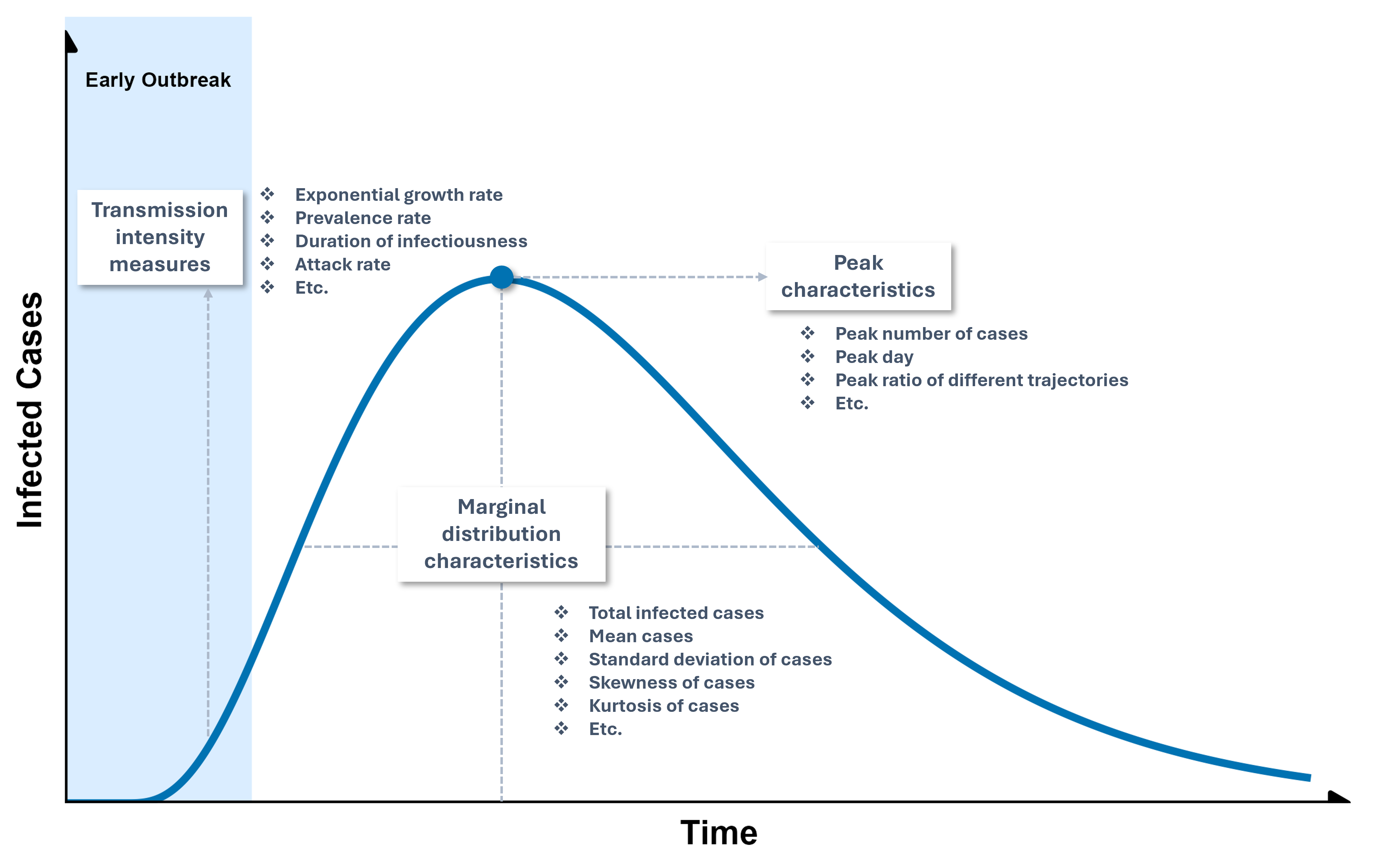}
    \caption{Conceptual illustration of epidemic cure features used to derive candidate summary statistics. Early outbreak transmission intensity measures characterize initial transmission dynamics, peak characteristics describe outbreak intensity and timing, and marginal distribution characteristics capture properties of the epidemic curve.}
    \label{figure:sumstats}
\end{figure}

Following the guidance from \citet{shearer2024opportunities}, we aim to select candidate statistics according to the epidemic phase and policy/analytical objectives. During early epidemic stages (Figure \ref{figure:sumstats}), it is important to measure transmission potential to characterize initial transmission dynamics, knowing peak characteristics and marginal distribution to capture properties of the epidemic curve. In later phases, when public health interventions are in place, behavioral and intervention-related statistics, such as the effective reproduction number, hospitalization rates, and incidence rate reductions, become crucial for evaluating intervention impacts and forecasting subsequent waves. This adaptive approach filters a manageable subset that is both statistically informative and relevant to practice, allowing efficient entropy-based selection while ensuring interpretability.

\subsection{Entropy minimization for statistics selection}
\label{s:entmin}

To select informative summary statistics in our methodological framework, we adopted the entropy-based subset selection method proposed by \citet{nunes2010optimal}. Specifically, we use the first stage of their two-stage procedure, which uses posterior entropy as a criterion for measuring the information content of different combinations of candidate summary statistics. The idea is that low entropy indicates a more concentrated and precise posterior, which contains more information content about the model parameters \citep{MR26286}. Specifically, for candidate subsets $s \subset \mathcal{S}$, we approximate the ABC posterior $\pi_{\text{ABC}}(\boldsymbol{\theta}|\boldsymbol{s}_{\text{obs}})$ and evaluate its differential entropy,
$$
H(\boldsymbol{\theta}|\boldsymbol{s}_{\text{obs}}) = -\int \pi_{\text{ABC}}(\boldsymbol{\theta}|\boldsymbol{s}_{\text{obs}}) \text{log} \pi_{\text{ABC}}(\boldsymbol{\theta}|\boldsymbol{s}_{\text{obs}}) d\boldsymbol{\theta}.
$$
We seek an optimal subset $\boldsymbol{s}^*$ that minimizes posterior entropy subject to a dimensionality constraint, $\boldsymbol{s}^*=\text{arg min}_{\boldsymbol{s} \subset \mathcal{S}}H(\boldsymbol{\theta}| \boldsymbol{s}_{\text{obs}}).$ The criterion favors summary statistics that jointly maximize information about $\boldsymbol{\theta}$ while penalizing redundancy. 

Following \citet{nunes2010optimal}, we estimate the entropy using the unbiased $k\text{th}$ nearest neighbor estimator of \citet{singh2003nearest} (suggested by \citet{singh2003nearest} and \citet{nunes2010optimal}, we use $k=4$ for its error properties); this estimator can be written as
$$
\hat{E}=\text{log}[\frac{\pi^{q/2}}{\Gamma(q/2+1)}]-\psi(k)+\text{log}n+\frac{q}{n}\sum_{i=1}^n \text{log}R_{i,k},
$$
where $q=\text{dim}(\boldsymbol{\theta})$, $\psi(k)=\Gamma'(k)/\Gamma(k)$ is the digamma function. Specifically, $R_{i,k}$ corresponds to the Euclidean distance from $\theta^i$ to its $k\text{th}$ nearest neighbor among the posterior sample $\{\theta^j\}_{j \neq i}$. This summation gives a score for how spread out or uncertain the posterior distribution is when using a certain set of summary statistics. Small values indicate that the data are clustered so that the majority of $k\text{th}$ nearest neighbor distances are small.

This entropy-based selection strategy offers several advantages for epidemic inference. First, the method provides an interpretable measure of inferential uncertainty by quantifying posterior precision through entropy, which is expected to inform public health decisions during early outbreaks when limited data and uncertainty quantification are common considerations. Second, the approach demonstrates flexibility with diverse summary statistics used in epidemic modeling. Finally, by penalizing redundancy, entropy minimization tends to select a minimally sufficient and exhaustive set of near-independent summary statistics, projecting the full continuous model trajectories into a discrete set of measurable.

\subsection{HMC for posterior sampling}
\label{s:hmc}

Following entropy-based summary statistic selection, we implement HMC for posterior sampling. The smooth, approximately Gaussian synthetic likelihood allows efficient gradient-based exploration of the posterior
$$
\pi(\boldsymbol{\theta}|\boldsymbol{s}^*_{\text{obs}}) \propto \mathcal{L}_{\text{SL}}(\boldsymbol{\theta})\pi({\boldsymbol{\theta}}),
$$
which is particularly advantageous for the high-dimensional, strongly correlated parameter spaces typical of compartmental models in epidemiology. The diagonal covariance structure further accelerates computation. HMC uses gradient information $\nabla_{\boldsymbol{\theta}} \text{log} \pi(\boldsymbol{\theta}|\boldsymbol{s}^*_{\text{obs}})$ to efficiently traverse the complex correlated posterior geometries. The synthetic log-likelihood decomposes across the $p$ summary statistics
$$
\text{log}\mathcal{L}_{\text{SL}}(\boldsymbol{\theta};\boldsymbol{s}^*_{\text{obs}})=-\frac{1}{2}\sum^p_{i=1}[\text{log}(2\pi\hat{\sigma}^2_i(\boldsymbol{\theta}))+\frac{(s^*_{i,\text{obs}}-\hat{\mu_i}(\boldsymbol{\theta}))^2}{\hat{\sigma}^2_i(\boldsymbol{\theta})}].
$$
Here, $\hat{\mu_i}(\boldsymbol{\theta})$ and $\hat{\sigma}^2_i(\boldsymbol{\theta})$ may be complex, nonlinear functions of the model parameters, as they are defined through repeated simulations of the model. Despite this complexity, the additive structure of the log-likelihood implies that gradient computations also decompose across summary statistics, enabling efficient automatic differentiation \citep{baydin2018automatic} within Stan, which provides an efficient implementation of HMC through NUTS \citep{duane1987hybrid,neal2012mcmc}. NUTS adaptively tunes path lengths and step sizes, alleviating the need for manual tuning and substantially improving robustness relative to traditional random-walk methods. By embedding our novel method in Stan, we make this Bayesian inference framework accessible to applied researchers.

\section{Simulation study}
\label{s:simulation}

We conducted a simulation study to evaluate the performance of the proposed methodological framework under controlled conditions where ground truth is known.

\subsection{Data and model}

We simulate epidemic data using a deterministic SEIR model for disease transmission in a closed population of size $N=10^4$. The model tracks individuals in four compartments: susceptible (S), exposed (E), infectious (I), and recovered (R), with dynamics governed by the following:
\begin{equation}
\begin{aligned}
\frac{dS(t)}{dt} &= -\frac{\beta I(t) S(t)}{N} \\
\frac{dE(t)}{dt} &= \frac{\beta I(t) S(t)}{N} - \sigma E(t) \\
\frac{dI(t)}{dt} &= \sigma E(t) - \gamma I(t) \\
\frac{dR(t)}{dt} &= \gamma I(t),
\end{aligned}
\end{equation}
where $\beta$ is the transmission rate, $\sigma$ is the rate of progression from exposed to infectious, and $\gamma$ is the recovery rate. The initial conditions were set to $S(0) = N -1, E(0)=0, I(0)=1,R(0)=0$. The true parameters were $\boldsymbol{\theta}_0=(\beta_0,\sigma_0, \gamma_0)=(0.4, 0.2, 0.07)$, corresponding to a basic reproduction number $R_0=\beta_0/\gamma_0\approx5.71$. The system was solved numerically using a fourth-order Runge-Kutta method \citep{butcher2016numerical} with daily time steps over a 100-day period. 

\subsection{Summary statistics selection}

From the trajectories of the exposed and infectious compartments, $(Y_E(t), Y_I(t))_{t=1}^{100}$, we constructed a candidate pool of 17 summary statistics that capture different aspects of epidemic dynamics. These include marginal distribution characteristics (mean, standard deviation, skewness, and kurtosis for both $Y_E(t)$ and $Y_I(t)$), peak characteristics (peak cases and timing for $Y_E(t)$ and $Y_I(t)$, plus the peak ratio), transmission intensity measures (duration above 50\% of peak infectiousness, early exponential growth rate, mean prevalence, and cumulative attack rate). By implementing entropy minimization ABC (as described in Section \ref{s:method}), we selected an optimal set that contained 4 summaries: standard deviation and peak prevalence for $Y_E(t)$, and kurtosis and peak cases for $Y_I(t)$.

\subsection{Results}

We specified weakly informative log-normal priors to ensure positivity while remaining broadly dispersed:
\begin{align*}
    \beta  &\sim \text{LogNormal}(\log(0.5),\, 0.5), \\
    \sigma &\sim \text{LogNormal}(\log(0.2),\, 0.5), \\
    \gamma &\sim \text{LogNormal}(\log(0.1),\, 0.5).
\end{align*}
These priors place substantial mass over plausible parameter scales while avoiding unrealistically large values.

Posterior sampling was performed using HMC via Stan. We ran 4 parallel chains for 2,000 iterations each, including the first 1,000 warm-up iterations used for sampler adaptation. Convergence was assessed using the rank-normalized split statistic $\hat{R}$ and the effective sample size (ESS) as implemented in Stan \citep{vehtari2021rank}. 

\begin{figure}[htbp]
\centering
\begin{subfigure}[b]{0.48\textwidth}
    \centering
    \includegraphics[width=\textwidth]{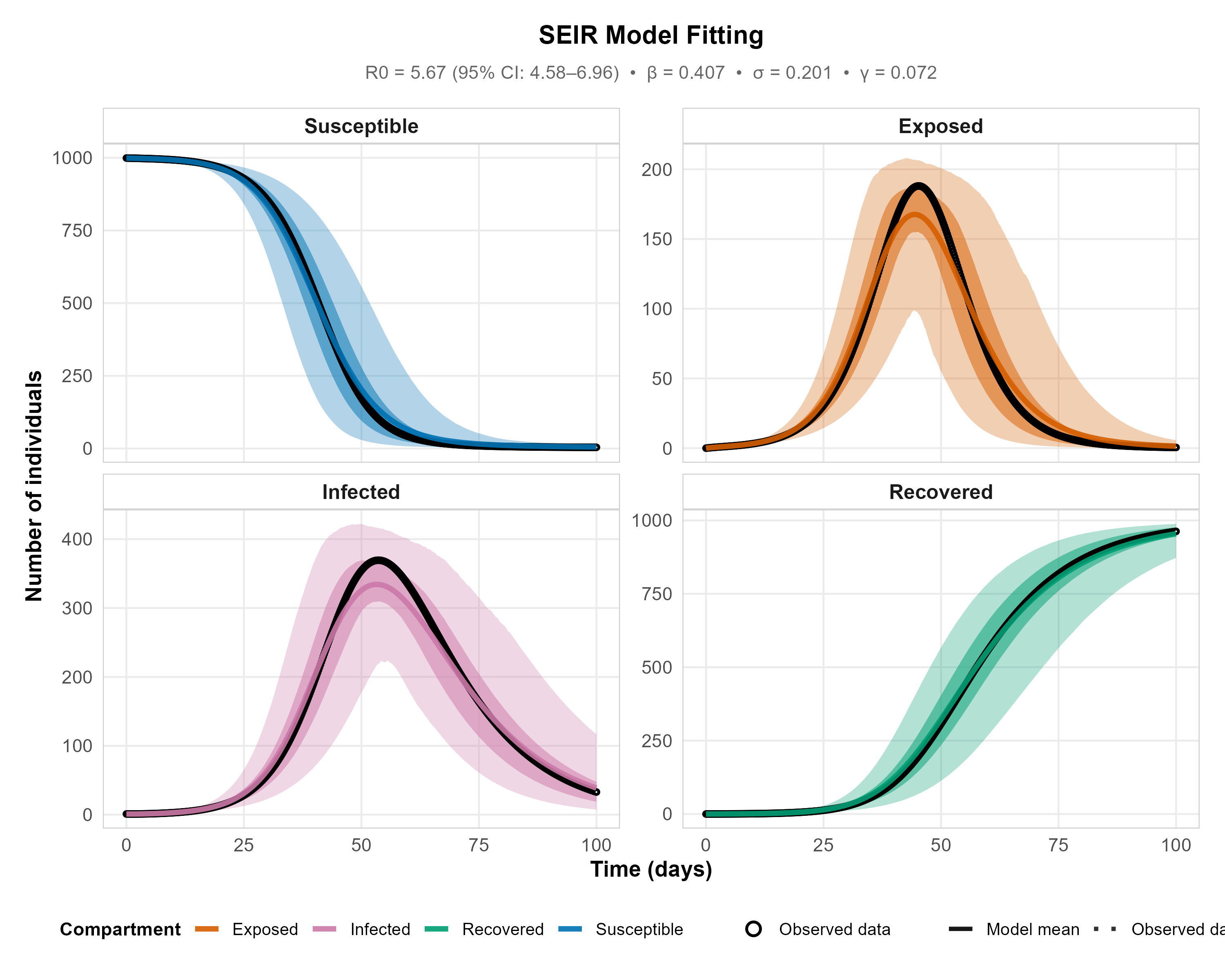}
    \caption{Posterior predictive trajectories}
    \label{fig:seir_fit}
\end{subfigure}
\hfill
\begin{subfigure}[b]{0.48\textwidth}
    \centering
    \includegraphics[width=\textwidth]{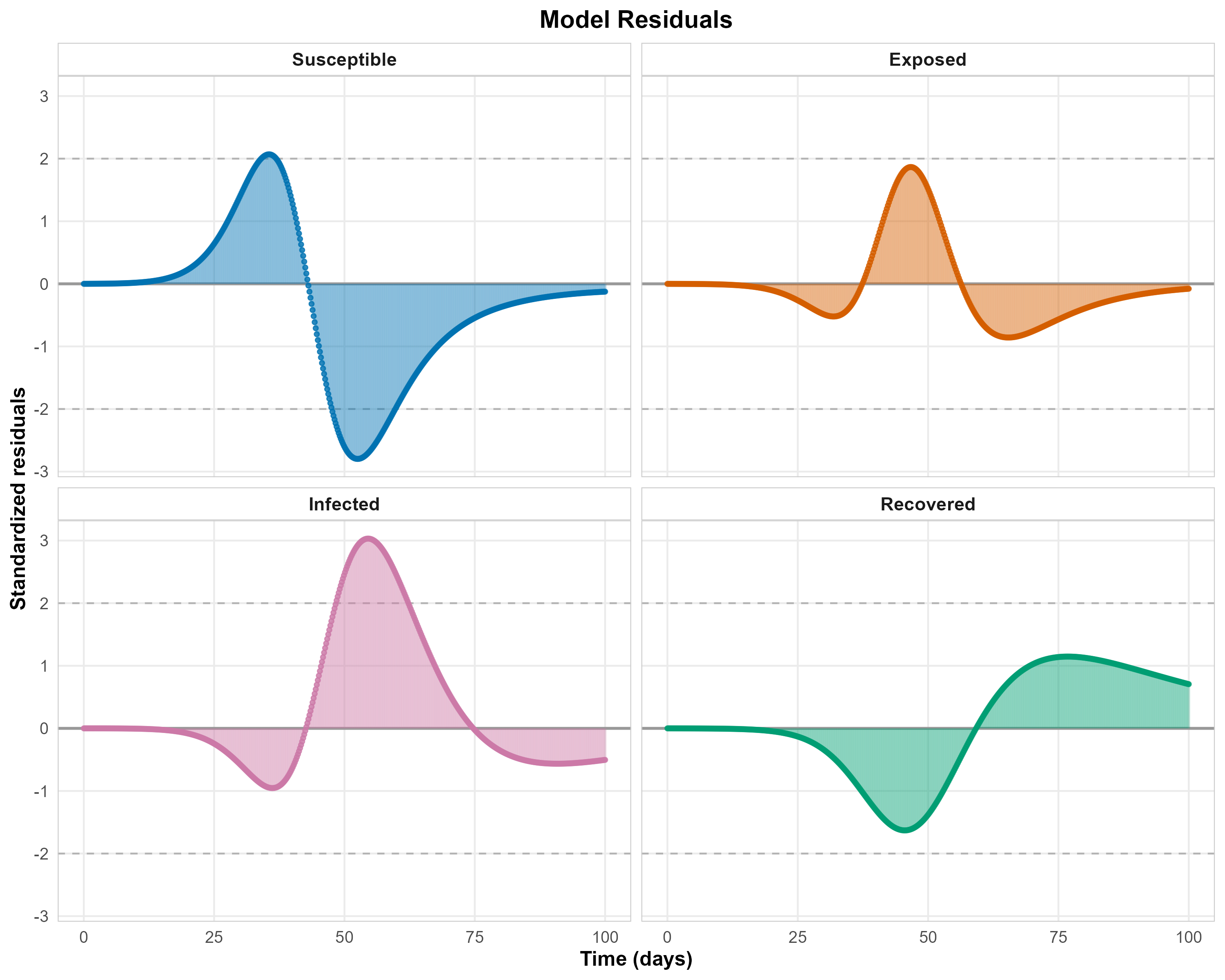}
    \caption{Standardized residuals by compartment}
    \label{fig:seir_residuals}
\end{subfigure}
\caption{SEIR model fitting and diagnostic assessment. 
(a) Posterior predictive trajectories for all four compartments, showing posterior mean and  95\% credible intervals (shaded regions) compared with observed data (black points). (b) Standardized residuals $(y_{obs} - \hat{y})/\text{SD}(\hat{y})$ by compartment, showing goodness-of-fit and systematic deviations across time. No systematic patterns or heteroscedasticity. The residuals oscillate around zero with approximately constant variance, supporting the adequacy of the synthetic likelihood approximation.}
\label{fig:seir_results}
\end{figure}

Figure \ref{fig:seir_fit} displays the posterior predictive trajectories for all four compartments, showing agreement between the model predictions and the observed data in all compartments. The shaded regions represent 95\% credible intervals, which provide good coverage of the observed data. Figure \ref{fig:seir_residuals} shows the standardized residuals by compartment, revealing no systematic patterns or trends that would suggest model misspecification. Table \ref{tab:convergence_diagnostics_simulation} shows that the posterior distributions exhibit small variances and 95\% credible intervals containing the true parameters, indicating that the parameters are estimated with high precision. All parameters obtained $\hat{R}<1.01$ and $\text{ESS}>1,000$, indicating good convergence and mixing. Additional model diagnostic results are provided in the Supplementary Material. 

\begin{table}[!htbp]
    \centering
    \caption{Convergence Diagnostics for Simulation Study}
    \label{tab:convergence_diagnostics_simulation}
    \resizebox{\ifdim\width>\linewidth\linewidth\else\width\fi}{!}{
    \begin{tabular}{@{}lccccccc@{}}
        \toprule
        Model          & Parameter    & True Value & Mean   & SD     & 95\% CI  & R-hat  & ESS  \\ 
        \midrule
        \multirow{5}{*}{SL Gamma} 
                        & \(\beta\)   & 0.400 & 0.407 & 0.049 & [0.307, 0.500]	 & 1.000 & 1563 \\
                        & \(\sigma\)  & 0.200 & 0.200 & 0.026 & [0.150, 0.253]	 & 1.001 & 1425 \\
                        & \(\gamma\)   & 0.070  & 0.072 & 0.011 & [0.051, 0.096] & 1.001 & 1955 \\
                        & \(R_0\) & 5.710 & 5.679 & 0.627 & [4.576, 6.999] & 1.000 & 4804 \\
        \cmidrule(lr){2-8}
                        & RMSE  & \multicolumn{4}{c}{12.568} \\
        \bottomrule
    \end{tabular}}
\end{table}

\section{Application: Influenza A outbreak}
\label{s:application}

We applied the proposed methodological framework to analyze the 1978 influenza A (H1N1) outbreak in a British boarding school \citep{center1978influenza, davies1982christ}. This real-world application demonstrates the use of our method for parameter inference when only limited surveillance data are available.

\subsection{Data and model}

The outbreak occurred among $N=763$ male boarding students in North England over a 14-day period, a total of 512 students were affected \citep{center1978influenza}. Data are publicly available in the R package \texttt{outbreaks} \citep{jombart2020outbreaks}.

We used a deterministic SIR model to describe the outbreak dynamics:
\begin{equation}
\begin{aligned}
\frac{dS}{dt} &= -\beta\frac{I(t)}{N}S(t) \\
\frac{dI}{dt} &= \beta\frac{I(t)}{N}S(t)-\gamma I(t) \\
\frac{dR}{dt} &= \gamma I(t),
\end{aligned}
\end{equation}
where $S(t), I(t), R(t)$ represents the number of susceptible, infectious and recovered individuals at time $t$, with the population size $N = S(t) + I(t) + R(t)$ remaining constant. The transmission rate $\beta$ governs the probability of infection, and the recovery rate $\gamma$ determines the mean infectious period $1/\gamma$. 

\citet{chatzilena2019contemporary} linked the deterministic model to the observed data using a Poisson observation model, $Y_t \sim \text{Poisson}(I(t))$, where $Y_t$ represents the daily count of confined students and $I(t)$ is the predicted number of cases. In Stan, discrete parameters cannot be directly sampled, so instead we used a Gamma observation model as a continuous approximation, parameterized as $Y_t \sim \text{Gamma}(I(t), 1)$ so that the mean and variance both equal $I(t)$, which avoids introducing discrete latent quantities while remaining compatible with HMC sampling.

We fixed $I(0) = 1$ and $R(0) = 0$, treating the initial fraction of susceptible $S_0$ as an estimable parameter to account for potential pre-existing immunity or uncertainty in the timing if the index case. We aim to estimate parameters $\boldsymbol{\theta} = (\beta, \gamma, S_0)$, from which we derive epidemiological interpretable quantities: basic reproduction number $R_0=\beta/\gamma$ and mean infectious period $1/\gamma$.

\subsection{Summary statistics selection}

We constructed a candidate pool of 10 summary statistics: (i) Marginal distribution characteristics (5 summaries): mean, standard deviation, skewness, kurtosis, and total number of infected individuals; (ii) Peak characteristics (2 summaries): peak cases and peak timing; (iii) Transmission intensity measures (3 summaries): duration greater than 10\% and 50\% of peak infectiousness, plus the early exponential growth rate. Applying the entropy minimization procedure, we get the optimal subset $S^*=\{\text{total infected, kurtosis, growth rate, peak cases}\}$.

\subsection{Results}

We specified weakly informative priors:
\begin{align*}
    \beta  &\sim \text{LogNormal}(\log(1.5),\, 0.5), \\
    \gamma &\sim \text{LogNormal}(\log(0.5),\, 0.5), \\
    s_0 &\sim \text{Beta}(50,\, 1).
\end{align*}
We sampled the posterior using HMC via Stan with 4 parallel chains of 2,000 iterations each (1,000 warm-up), producing 4,000 posterior samples. For validation, we compared our approach to the Poisson likelihood used by \citet{chatzilena2019contemporary} without using summary statistics.

Table \ref{tab:convergence_diagnostics_application} reports the posterior summaries and computational metrics for both methods. The BSL estimates are closely aligned with the Poisson method of reference. As expected, the BSL credible intervals are wider than the benchmark. While this partly reflects dimension reduction to four summary statistics, the narrower benchmark intervals appear to understate posterior uncertainty. In contrast, the BSL intervals exhibit more appropriate uncertainty quantification, with $R_0$ estimated within $\pm0.3$ reproductive units. 

\begin{figure}[htbp]
\centering
\begin{subfigure}[b]{0.48\textwidth}
    \centering
    \includegraphics[width=\textwidth]{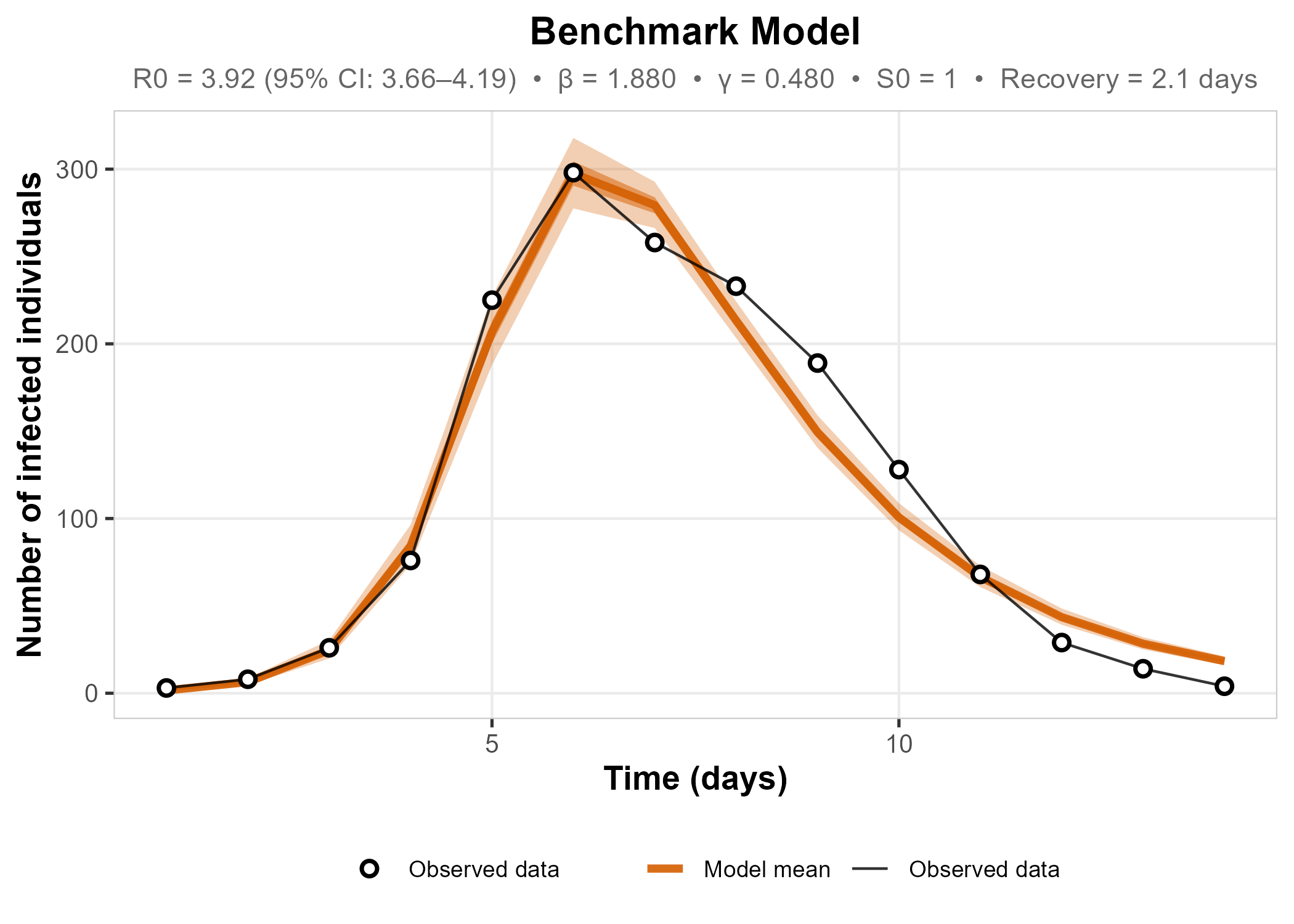}
    \caption{Benchmark model fitted to observed data}
    \label{fig:sir_benchmark}
\end{subfigure}
\hfill
\begin{subfigure}[b]{0.48\textwidth}
    \centering
    \includegraphics[width=\textwidth]{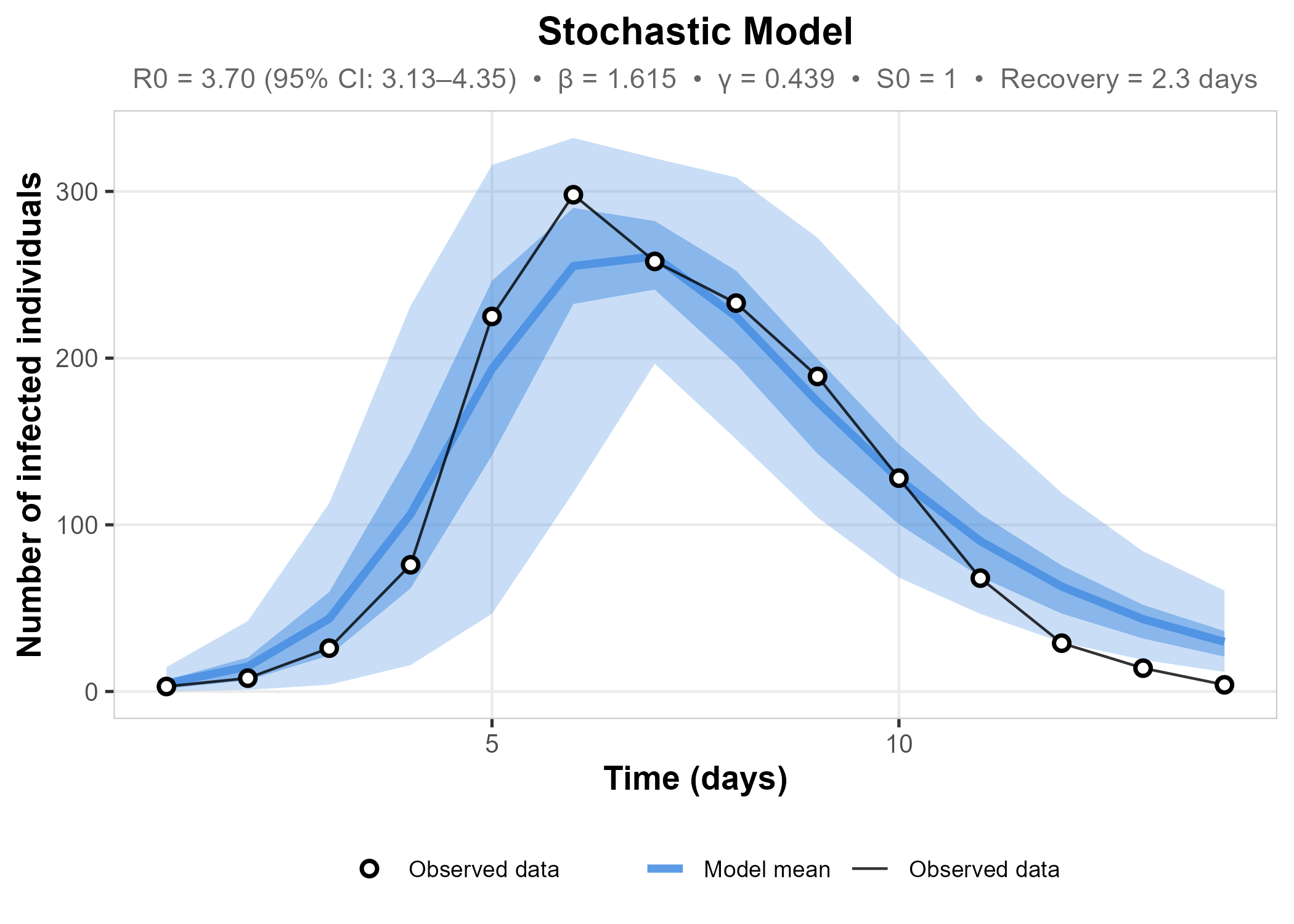}
    \caption{Synthetic likelihood model fitted to observed data}
    \label{fig:sir_synlik}
\end{subfigure}
\caption{Model fit comparison for influenza A outbreak study. 
(a) The benchmark model provides a deterministic fit to observed data. (b) The synthetic likelihood model incorporates more uncertainty. Black points represent observed data, lines show model means, and shared regions indicate 95\% credible intervals.}
\label{fig:sir_results}
\end{figure}

Figure \ref{fig:sir_results} displays posterior predictive fits to the observed data for both methods. The benchmark model (Figure \ref{fig:sir_benchmark}) provides a tight deterministic fit, with narrow credible intervals, whereas the BSL model (Figure \ref{fig:sir_synlik}) captures the epidemic trajectory with wider but appropriate uncertainty quantification: observed counts fall within the 95\% posterior predictive intervals for 12 of 14 days, with only slight overestimation on the final two days. In comparison, the benchmark only captures 7 of 14 days. Both models reproduce the rapid exponential rise, peak timing, and subsequent decay phase. Additional diagnostic results are provided in the Supplementary Material. With $R_0 = 3.70$, the outbreak doubling time during the exponential growth phase was approximately 1/8 days, consistent with the observed rapid increase in cases.

\begin{table}[!htbp]
    \caption{Convergence Diagnostics with Computation Metrics for Application Study}
    \label{tab:convergence_diagnostics_application}
    \resizebox{\ifdim\width>\linewidth\linewidth\else\width\fi}{!}{
    \begin{tabular}{@{}lcccccc@{}}
        \toprule
        Model          & Parameter    & Mean   & SD     & 95\% CI  & R-hat  & ESS  \\ 
        \midrule
        \multirow{5}{*}{Poisson}      
                        & \(\beta\)   & 1.881 & 0.054 & [1.794, 1.972] & 1.004 & 1301 \\
                        & \(\gamma\)  & 0.479 & 0.011 & [0.462, 0.499] & 1.002 & 1692 \\
                        & \(S_0\)     & 0.999 & 0.000 & [0.999, 1.000] & 1.004 & 1353\\
                        & \(R_0\)& 3.917 & 0.141 & [3.697, 4.158] & 1.002 & 1727 \\
                        & \(1/\gamma\)    & 2.083 & 0.049 & [2.005, 2.165]	 & 1.002 & 1692 \\
        \cmidrule(lr){2-7}
                        & Time (sec)  & \multicolumn{4}{c}{10.1} \\
                        & RMSE  & \multicolumn{4}{c}{17.303} \\
        \midrule
        \multirow{5}{*}{SL Gamma} 
                        & \(\beta\)   & 1.618 & 0.101 & [1.463, 1.791]	 & 1.001 & 1614 \\
                        & \(\gamma\)  & 0.439 & 0.032 & [0.388, 0.494]	 & 1.000 & 1837  \\
                        & \(S_0\)     & 0.998 & 0.002 & [0.995, 1.000] & 1.014 & 344 \\
                        & \(R_0\)& 3.701 & 0.314 & [3.211, 4.235] & 1.001 & 2337 \\
                        & \(1/\gamma\)    & 2.289 & 0.167 & [2.023, 2.574]	 & 1.000 & 1837 \\
        \cmidrule(lr){2-7}
                        & Time (sec)  & \multicolumn{4}{c}{27.3} \\
                        & RMSE  & \multicolumn{4}{c}{24.098} \\
        \bottomrule
    \end{tabular}}
\end{table}

\section{Discussion and conclusion}
\label{s:discuss}

In this study, we propose a hybrid Bayesian inference framework for compartmental models in epidemiology that integrates policy-driven summary statistics construction, entropy-based subset selection, synthetic likelihoods with simplified covariance structure, and HMC sampling in Stan. Through studies of simulated SEIR systems and analysis of the 1978 influenza A outbreak, we demonstrate that carefully chosen low-dimensional set of summary statistics perform well in parameter estimation while preserving uncertainty quantification. The entropy-based selection procedure identifies compact subsets of four summary statistics from candidate pools of 10-17, achieving dimension reductions of 60-76\% while preserving information about transmission dynamics, epidemic timing, and population-level impact. 

The framework offers several advantages. First, it grounds statistical methodology in public health practice, selected statistics have clear policy interpretation rather than being arbitrary mathematical constructs. This alignment improves interpretability and ensures that inference directly targets quantities relevant to intervention planning. Second, entropy minimization provides objective guidance for choosing among correlated policy-relevant quantities, producing a minimally sufficient and exhaustive set of near-independent summary statistics that forms a basis spanning the posterior space, thereby justifiably simplifying the covariance structure. Third, the integrated framework delivers both computational efficiency and inferential accuracy on metrics that matter, including parameter coverage and predictive performance for policy-relevant quantities. Most importantly, this integrated approach, ABC entropy minimization fed into simplified BSL sampled via HMC, yields practical gains over application of individual methods, thereby bridging the gap between likelihood-free methods and likelihood-based Bayesian inference.

For epidemiologists, these advantages have direct practical implications for outbreak investigation and decision-making. Most importantly, our method provides uncertainty quantification that better reflects real-world inferential limitations. The wider credible intervals produced by the proposed approach compared to the benchmark methods present uncertainty more realistically  when working with limited surveillance data and model approximations (Figure \ref{fig:sir_results}). This protects against overconfident policy recommendations. Furthermore, by implementing synthetic likelihood in Stan, we make the novel Bayesian inference method accessible to applied researchers, allowing epidemiologists to fit models without tractable likelihood functions in Stan.

As with all approximate methods, the proposed approach has several methodological trade-offs. A key limitation lies in the computational burden of subset selection when the candidate pool of summary statistics becomes large. Exhaustive enumeration of all possible subsets $2^p-1$ grows exponentially with the number of candidates $\boldsymbol{s}=(s_1,...,s_p)$ and becomes infeasible beyond moderate dimensions, particularly in settings involving temporal patterns, spatial spread, age structure, severity, and intervention effects. Our policy-driven, phase specific pre-selection mitigates this challenge, and scalable selection algorithms is important for high-dimensional problems. In addition, the BSL posterior involves multiple layers of approximation, including Gaussian assumptions for the distribution of summary statistics and estimation of their moments from finite simulations. Unlike previous approaches that impose independence heuristically or rely on ad hoc dimension reduction, our framework is based on the well-documented theoretical properties of entropy-selection, which promote weak dependence among selected summaries. This provides a justification for adopting a diagonal covariance structure as a tractable approximation. Empirically, this assumption performs well; however, performance may be limited when summaries exhibit strong non-Gaussian behavior. 

Looking forward, several extensions are promising. Alternative subset selection methods, incorporating mutual information criteria, projection techniques, or machine learning-based summary construction, could extend our framework to high-dimensional candidate spaces \citep{prangle2018summary, ahmadi2025machine}. More flexible covariance representations, such as shrinkage estimators, factor models, or structured covariance matrices \citep{ong2018likelihood, an2019accelerating, priddle2022efficient}, could relax the independence assumptions and be computationally efficient. Computational efficiency could also be improved by using surrogate modeling of summary statistics, such as Gaussian process emulators \citep{gramacy2020surrogates}, reducing the need for repeated simulations. More broadly, application beyond infectious disease to ecological \citep{komorowski2009bayesian}, biological systems \citep{fearnhead2014inference}, phylogenetic \citep{king2025exact} would further demonstrate its generality and potentially inspire domain-specific innovations.

In conclusion, this work contributes to the ongoing development of Bayesian inference methods for epidemiological modeling by integrating the strengths of ABC, BSL and HMC within a framework. By addressing persistent challenges in likelihood-free inference, namely summary statistics selection and efficient posterior exploration, the proposed approach offers a practical solution for compartmental models. While the proposed framework demonstrates promise empirically, its full potential will depend on further validation and extension to more complex cases. The code implementing our proposed method is freely available at \url{https://github.com/XiahuiLi-ab/HybridBayes.git}.


\section*{Acknowledgments}

XL acknowledges funding from the University of St Andrews and the China Scholarship Council.\vspace*{-8pt}

\section*{Supplementary Materials}

Supplementary material referenced in Section~\ref{s:simulation} and Section~\ref{s:application}, is available with
this article on the Biometrics website in the Wiley Online
Library.\vspace*{-8pt}

\bibliographystyle{vancouver}
\bibliography{main}

@article{kermack1927contribution,
  title={A contribution to the mathematical theory of epidemics},
  author={Kermack, William Ogilvy and McKendrick, Anderson G},
  journal={Proceedings of the royal society of london. Series A, Containing papers of a mathematical and physical character},
  volume={115},
  number={772},
  pages={700--721},
  year={1927},
  publisher={The Royal Society London}
}

@article{komorowski2009bayesian,
  title={Bayesian inference of biochemical kinetic parameters using the linear noise approximation},
  author={Komorowski, Micha{\l} and Finkenst{\"a}dt, B{\"a}rbel and Harper, Claire V and Rand, David A},
  journal={BMC bioinformatics},
  volume={10},
  number={1},
  pages={343},
  year={2009},
  publisher={Springer}
}

@article{fearnhead2014inference,
  title={Inference for reaction networks using the linear noise approximation},
  author={Fearnhead, Paul and Giagos, Vasilieos and Sherlock, Chris},
  journal={Biometrics},
  volume={70},
  number={2},
  pages={457--466},
  year={2014},
  publisher={Oxford University Press}
}

@article{chatzilena2019contemporary,
  title={Contemporary statistical inference for infectious disease models using Stan},
  author={Chatzilena, Anastasia and van Leeuwen, Edwin and Ratmann, Oliver and Baguelin, Marc and Demiris, Nikolaos},
  journal={Epidemics},
  volume={29},
  pages={100367},
  year={2019},
  publisher={Elsevier}
}

@article{nunes2010optimal,
  title={On optimal selection of summary statistics for approximate Bayesian computation.},
  author={Nunes, Matthew A and Balding, David J},
  journal={Statistical Applications in Genetics \& Molecular Biology},
  volume={9},
  number={1},
  year={2010}
}

@article {MR26286,
    AUTHOR = {Shannon, C. E.},
     TITLE = {A mathematical theory of communication},
   JOURNAL = {Bell System Tech. J.},
  FJOURNAL = {The Bell System Technical Journal},
    VOLUME = {27},
      YEAR = {1948},
     PAGES = {379--423, 623--656},
      ISSN = {0005-8580},
   MRCLASS = {60.0X},
  MRNUMBER = {26286},
MRREVIEWER = {J.\ L.\ Doob},
       DOI = {10.1002/j.1538-7305.1948.tb01338.x},
       URL = {https://doi.org/10.1002/j.1538-7305.1948.tb01338.x},
}

@article{singh2003nearest,
  title={Nearest neighbor estimates of entropy},
  author={Singh, Harshinder and Misra, Neeraj and Hnizdo, Vladimir and Fedorowicz, Adam and Demchuk, Eugene},
  journal={American journal of mathematical and management sciences},
  volume={23},
  number={3-4},
  pages={301--321},
  year={2003},
  publisher={Taylor \& Francis}
}

@article{carpenter2017stan,
  title={Stan: A probabilistic programming language},
  author={Carpenter, Bob and Gelman, Andrew and Hoffman, Matthew D and Lee, Daniel and Goodrich, Ben and Betancourt, Michael and Brubaker, Marcus A and Guo, Jiqiang and Li, Peter and Riddell, Allen},
  journal={Journal of statistical software},
  volume={76},
  year={2017},
  publisher={NIH Public Access}
}

@article{duane1987hybrid,
  title={Hybrid monte carlo},
  author={Duane, Simon and Kennedy, Anthony D and Pendleton, Brian J and Roweth, Duncan},
  journal={Physics letters B},
  volume={195},
  number={2},
  pages={216--222},
  year={1987},
  publisher={Elsevier}
}

@article{neal2012mcmc,
  title={MCMC using Hamiltonian dynamics},
  author={Neal, Radford M},
  journal={arXiv preprint arXiv:1206.1901},
  year={2012}
}

@article{baydin2018automatic,
  title={Automatic differentiation in machine learning: a survey},
  author={Baydin, Atilim Gunes and Pearlmutter, Barak A and Radul, Alexey Andreyevich and Siskind, Jeffrey Mark},
  journal={Journal of machine learning research},
  volume={18},
  number={153},
  pages={1--43},
  year={2018}
}

@article{brauer2008compartmental,
  title={Compartmental models in epidemiology},
  author={Brauer, Fred},
  journal={Mathematical epidemiology},
  pages={19--79},
  year={2008},
  publisher={Springer}
}

@article{swallow2022challenges,
  title={Challenges in estimation, uncertainty quantification and elicitation for pandemic modelling},
  author={Swallow, Ben and Birrell, Paul and Blake, Joshua and Burgman, Mark and Challenor, Peter and Coffeng, Luc E and Dawid, Philip and De Angelis, Daniela and Goldstein, Michael and Hemming, Victoria and others},
  journal={Epidemics},
  volume={38},
  pages={100547},
  year={2022},
  publisher={Elsevier}
}

@article{kypraios2017tutorial,
  title={A tutorial introduction to Bayesian inference for stochastic epidemic models using Approximate Bayesian Computation},
  author={Kypraios, Theodore and Neal, Peter and Prangle, Dennis},
  journal={Mathematical biosciences},
  volume={287},
  pages={42--53},
  year={2017},
  publisher={Elsevier}
}

@article{grinsztajn2021bayesian,
  title={Bayesian workflow for disease transmission modeling in Stan},
  author={Grinsztajn, L{\'e}o and Semenova, Elizaveta and Margossian, Charles C and Riou, Julien},
  journal={Statistics in medicine},
  volume={40},
  number={27},
  pages={6209--6234},
  year={2021},
  publisher={Wiley Online Library}
}

@book{brooks2011handbook,
  title={Handbook of markov chain monte carlo},
  author={Brooks, Steve and Gelman, Andrew and Jones, Galin and Meng, Xiao-Li},
  year={2011},
  publisher={CRC press}
}

@article{metropolis1953equation,
  title={Equation of state calculations by fast computing machines},
  author={Metropolis, Nicholas and Rosenbluth, Arianna W and Rosenbluth, Marshall N and Teller, Augusta H and Teller, Edward},
  journal={The journal of chemical physics},
  volume={21},
  number={6},
  pages={1087--1092},
  year={1953},
  publisher={American Institute of Physics}
}

@article{hastings1970monte,
  title={Monte Carlo sampling methods using Markov chains and their applications},
  author={Hastings, W Keith},
  year={1970},
  journal={Biometrika},
  publisher={Oxford University Press}
}

@article{hoffman2014no,
  title={The No-U-Turn sampler: adaptively setting path lengths in Hamiltonian Monte Carlo.},
  author={Hoffman, Matthew D and Gelman, Andrew and others},
  journal={J. Mach. Learn. Res.},
  volume={15},
  number={1},
  pages={1593--1623},
  year={2014}
}

@article{beaumont2019approximate,
  title={Approximate bayesian computation},
  author={Beaumont, Mark A},
  journal={Annual review of statistics and its application},
  volume={6},
  number={1},
  pages={379--403},
  year={2019},
  publisher={Annual Reviews}
}

@incollection{sisson2018overview,
  title={Overview of ABC},
  author={Sisson, Scott A and Fan, Yanan and Beaumont, Mark A},
  booktitle={Handbook of approximate Bayesian computation},
  pages={3--54},
  year={2018},
  publisher={Chapman and Hall/CRC}
}

@incollection{prangle2018summary,
  title={Summary statistics},
  author={Prangle, Dennis},
  booktitle={Handbook of approximate Bayesian computation},
  pages={125--152},
  year={2018},
  publisher={Chapman and Hall/CRC}
}

@article{blum2010approximate,
  title={Approximate Bayesian computation: a nonparametric perspective},
  author={Blum, Michael GB},
  journal={Journal of the American Statistical Association},
  volume={105},
  number={491},
  pages={1178--1187},
  year={2010},
  publisher={Taylor \& Francis}
}

@article{toni2009approximate,
  title={Approximate Bayesian computation scheme for parameter inference and model selection in dynamical systems},
  author={Toni, Tina and Welch, David and Strelkowa, Natalja and Ipsen, Andreas and Stumpf, Michael PH},
  journal={Journal of the Royal Society Interface},
  volume={6},
  number={31},
  pages={187--202},
  year={2009},
  publisher={The Royal Society London}
}

@inproceedings{jiang2018approximate,
  title={Approximate Bayesian computation with Kullback-Leibler divergence as data discrepancy},
  author={Jiang, Bai},
  booktitle={International conference on artificial intelligence and statistics},
  pages={1711--1721},
  year={2018},
  organization={PMLR}
}

@article{retkute2025novel,
  title={A novel two-stage parameter estimation framework integrating Approximate Bayesian Computation and Machine Learning: The ABC-RF-rejection algorithm},
  author={Retkute, Renata and Gilligan, Christopher A},
  journal={arXiv preprint arXiv:2507.02072},
  year={2025}
}

@article{shearer2024opportunities,
  title={Opportunities to strengthen respiratory virus surveillance systems in Australia: lessons learned from the COVID-19 response},
  author={Shearer, Freya M and Edwards, Laura and Kirk, Martyn and Eales, Oliver and Golding, Nick and Hassall, Jenna and Liu, Bette and Lydeamore, Michael and Miller, Caroline and Moss, Robert and others},
  journal={Communicable Diseases Intelligence},
  volume={48},
  year={2024},
  publisher={Australian Government Department of Health and Ageing}
}

@article{wood2010statistical,
  title={Statistical inference for noisy nonlinear ecological dynamic systems},
  author={Wood, Simon N},
  journal={Nature},
  volume={466},
  number={7310},
  pages={1102--1104},
  year={2010},
  publisher={Nature Publishing Group UK London}
}

@article{price2018bayesian,
  title={Bayesian synthetic likelihood},
  author={Price, Leah F and Drovandi, Christopher C and Lee, Anthony and Nott, David J},
  journal={Journal of Computational and Graphical Statistics},
  volume={27},
  number={1},
  pages={1--11},
  year={2018},
  publisher={Taylor \& Francis}
}

@article{an2019accelerating,
  title={Accelerating Bayesian synthetic likelihood with the graphical lasso},
  author={An, Ziwen and South, Leah F and Nott, David J and Drovandi, Christopher C},
  journal={Journal of Computational and Graphical Statistics},
  volume={28},
  number={2},
  pages={471--475},
  year={2019},
  publisher={Taylor \& Francis}
}

@article{ong2018likelihood,
  title={Likelihood-free inference in high dimensions with synthetic likelihood},
  author={Ong, Victor M-H and Nott, David J and Tran, Minh-Ngoc and Sisson, Scott A and Drovandi, Christopher C},
  journal={Computational Statistics \& Data Analysis},
  volume={128},
  pages={271--291},
  year={2018},
  publisher={Elsevier}
}

@article{priddle2022efficient,
  title={Efficient Bayesian synthetic likelihood with whitening transformations},
  author={Priddle, Jacob W and Sisson, Scott A and Frazier, David T and Turner, Ian and Drovandi, Christopher},
  journal={Journal of Computational and Graphical Statistics},
  volume={31},
  number={1},
  pages={50--63},
  year={2022},
  publisher={Taylor \& Francis}
}

@article{frazier2023bayesian,
  title={Bayesian inference using synthetic likelihood: asymptotics and adjustments},
  author={Frazier, David T and Nott, David J and Drovandi, Christopher and Kohn, Robert},
  journal={Journal of the American Statistical Association},
  volume={118},
  number={544},
  pages={2821--2832},
  year={2023},
  publisher={Taylor \& Francis}
}

@article{osthus2017forecasting,
  title={Forecasting seasonal influenza with a state-space SIR model},
  author={Osthus, Dave and Hickmann, Kyle S and Caragea, Petru{\c{t}}a C and Higdon, Dave and Del Valle, Sara Y},
  journal={The annals of applied statistics},
  volume={11},
  number={1},
  pages={202},
  year={2017}
}

@article{fasiolo2018extended,
  title={An extended empirical saddlepoint approximation for intractable likelihoods},
  author={Fasiolo, Matteo and Wood, Simon N and Hartig, Florian and Bravington, Mark V},
  year={2018},
  journal={Electronic Journal of Statistics}
}

@article{he2013inferring,
  title={Inferring the causes of the three waves of the 1918 influenza pandemic in England and Wales},
  author={He, Daihai and Dushoff, Jonathan and Day, Troy and Ma, Junling and Earn, David JD},
  journal={Proceedings of the Royal Society B: Biological Sciences},
  volume={280},
  number={1766},
  pages={20131345},
  year={2013},
  publisher={The Royal Society}
}

@article{camacho2011explaining,
  title={Explaining rapid reinfections in multiple-wave influenza outbreaks: Tristan da Cunha 1971 epidemic as a case study},
  author={Camacho, Anton and Ballesteros, S{\'e}bastien and Graham, Andrea L and Carrat, Fabrice and Ratmann, Oliver and Cazelles, Bernard},
  journal={Proceedings of the Royal Society B: Biological Sciences},
  volume={278},
  number={1725},
  pages={3635--3643},
  year={2011},
  publisher={The Royal Society}
}

@article{heesterbeek2015modeling,
  title={Modeling infectious disease dynamics in the complex landscape of global health},
  author={Heesterbeek, Hans and Anderson, Roy M and Andreasen, Viggo and Bansal, Shweta and De Angelis, Daniela and Dye, Chris and Eames, Ken TD and Edmunds, W John and Frost, Simon DW and Funk, Sebastian and others},
  journal={Science},
  volume={347},
  number={6227},
  pages={aaa4339},
  year={2015},
  publisher={American Association for the Advancement of Science}
}

@book{butcher2016numerical,
  title={Numerical methods for ordinary differential equations},
  author={Butcher, John Charles},
  year={2016},
  publisher={John Wiley \& Sons}
}

@article{ahmadi2025machine,
  title={Machine learned summary statistics for Bayesian inference of systems biology models parameters: opportunities and challenges},
  author={Ahmadi, Atiyeh and Podina, Lena and H{\"o}pfl, Sebastian and Ingalls, Brian},
  journal={Current Opinion in Systems Biology},
  pages={100560},
  year={2025},
  publisher={Elsevier}
}

@article{vehtari2021rank,
  title={Rank-normalization, folding, and localization: An improved R-hat for assessing convergence of MCMC (with discussion)},
  author={Vehtari, Aki and Gelman, Andrew and Simpson, Daniel and Carpenter, Bob and B{\"u}rkner, Paul-Christian},
  journal={Bayesian analysis},
  volume={16},
  number={2},
  pages={667--718},
  year={2021},
  publisher={International Society for Bayesian Analysis}
}

@article{wilkinson2009stochastic,
  title={Stochastic modelling for quantitative description of heterogeneous biological systems},
  author={Wilkinson, Darren J},
  journal={Nature Reviews Genetics},
  volume={10},
  number={2},
  pages={122--133},
  year={2009},
  publisher={Nature Publishing Group UK London}
}

@article{li2025advances,
  title={Advances in approximate Bayesian inference for models in epidemiology},
  author={Li, Xiahui and Chadwick, Fergus and Swallow, Ben},
  journal={Epidemics},
  pages={100855},
  year={2025},
  publisher={Elsevier}
}

@article{o2010introduction,
  title={Introduction and snapshot review: relating infectious disease transmission models to data},
  author={O'Neill, Philip D},
  journal={Statistics in medicine},
  volume={29},
  number={20},
  pages={2069--2077},
  year={2010},
  publisher={Wiley Online Library}
}

@book{gramacy2020surrogates,
  title={Surrogates: Gaussian process modeling, design, and optimization for the applied sciences},
  author={Gramacy, Robert B},
  year={2020},
  publisher={Chapman and Hall/CRC}
}

@article{king2025exact,
  title={Exact phylodynamic likelihood via structured Markov genealogy processes},
  author={King, Aaron A and Lin, Qianying and Ionides, Edward L},
  journal={ArXiv},
  pages={arXiv--2405},
  year={2025}
}

@misc{jombart2020outbreaks,
  author={Jombart, T and Frost, S and Nouvellet, P and Campbell, F and Sudre, B},
  title={Outbreaks: a collection of disease outbreak data. R package version 1.9. 0},
  year={2020},
  howpublished={See https://CRAN. R-project. org/package= outbreaks}
}

@article{center1978influenza,
  title={Influenza in a boarding school},
  author={Center, CDS},
  journal={BMJ},
  volume={1},
  pages={587},
  year={1978}
}

@article{davies1982christ,
  title={Christ's Hospital 1978--79: An account of two outbreaks of influenza A H1N1},
  author={Davies, JR and Smith, AJ and Grilli, EA and Hoskins, TW},
  journal={Journal of Infection},
  volume={5},
  number={2},
  pages={151--156},
  year={1982},
  publisher={Elsevier}
}

@article{berger2021rational,
  title={Rational policymaking during a pandemic},
  author={Berger, Lo{\"\i}c and Berger, Nicolas and Bosetti, Valentina and Gilboa, Itzhak and Hansen, Lars Peter and Jarvis, Christopher and Marinacci, Massimo and Smith, Richard D},
  journal={Proceedings of the National Academy of Sciences},
  volume={118},
  number={4},
  pages={e2012704118},
  year={2021},
  publisher={National Academy of Sciences}
}

@article{shearer2020infectious,
  title={Infectious disease pandemic planning and response: Incorporating decision analysis},
  author={Shearer, Freya M and Moss, Robert and McVernon, Jodie and Ross, Joshua V and McCaw, James M},
  journal={PLoS medicine},
  volume={17},
  number={1},
  pages={e1003018},
  year={2020},
  publisher={Public Library of Science San Francisco, CA USA}
}

@article{cauchemez2008estimating,
  title={Estimating the impact of school closure on influenza transmission from Sentinel data},
  author={Cauchemez, Simon and Valleron, Alain-Jacques and Boelle, Pierre-Yves and Flahault, Antoine and Ferguson, Neil M},
  journal={Nature},
  volume={452},
  number={7188},
  pages={750--754},
  year={2008},
  publisher={Nature Publishing Group UK London}
}

@article{desai2019real,
  title={Real-time epidemic forecasting: challenges and opportunities},
  author={Desai, Angel N and Kraemer, Moritz UG and Bhatia, Sangeeta and Cori, Anne and Nouvellet, Pierre and Herringer, Mark and Cohn, Emily L and Carrion, Malwina and Brownstein, John S and Madoff, Lawrence C and others},
  journal={Health security},
  volume={17},
  number={4},
  pages={268--275},
  year={2019},
  publisher={Mary Ann Liebert, Inc., publishers 140 Huguenot Street, 3rd Floor New~…}
}

@article{fischer2016cdc,
  title={CDC grand rounds: modeling and public health decision-making},
  author={Fischer, Leah S},
  journal={MMWR. Morbidity and Mortality Weekly Report},
  volume={65},
  year={2016}
}

@article{thompson2026infectious,
  title={Infectious disease outbreak controllability: biological, social and public health factors},
  author={Thompson, Robin N and Bansal, Shweta and Clapham, Hannah and Dyson, Louise and Gutierrez, Maria A and Hadley, Liza and Hart, William S and Heesterbeek, Hans and Hollingsworth, T Deirdre and House, Thomas and others},
  journal={Proceedings of the Royal Society B: Biological Sciences},
  volume={293},
  number={2063},
  year={2026},
  publisher={The Royal Society}
}

@article{grassly2008mathematical,
  title={Mathematical models of infectious disease transmission},
  author={Grassly, Nicholas C and Fraser, Christophe},
  journal={Nature Reviews Microbiology},
  volume={6},
  number={6},
  pages={477--487},
  year={2008},
  publisher={Nature Publishing Group UK London}
}

@techreport{world2000report,
  title={WHO report on global surveillance of epidemic-prone infectious diseases},
  author={World Health Organization and others},
  year={2000},
  institution={World Health Organization}
}

@article{moran2016epidemic,
  title={Epidemic forecasting is messier than weather forecasting: The role of human behavior and internet data streams in epidemic forecast},
  author={Moran, Kelly R and Fairchild, Geoffrey and Generous, Nicholas and Hickmann, Kyle and Osthus, Dave and Priedhorsky, Reid and Hyman, James and Del Valle, Sara Y},
  journal={The Journal of infectious diseases},
  volume={214},
  number={suppl\_4},
  pages={S404--S408},
  year={2016},
  publisher={Oxford University Press}
}

@article{balabdaoui2020age,
  title={Age-stratified discrete compartment model of the COVID-19 epidemic with application to Switzerland},
  author={Balabdaoui, Fadoua and Mohr, Dirk},
  journal={Scientific reports},
  volume={10},
  number={1},
  pages={21306},
  year={2020},
  publisher={Nature Publishing Group UK London}
}

@article{chang2021mobility,
  title={Mobility network models of COVID-19 explain inequities and inform reopening},
  author={Chang, Serina and Pierson, Emma and Koh, Pang Wei and Gerardin, Jaline and Redbird, Beth and Grusky, David and Leskovec, Jure},
  journal={Nature},
  volume={589},
  number={7840},
  pages={82--87},
  year={2021},
  publisher={Nature Publishing Group UK London}
}

\appendix

\label{lastpage}

\end{document}